\documentclass[12pt]{article}
\usepackage{graphicx}
\begin{document}

\title{Mean field approximation for noisy delay coupled excitable neurons }
 \author{
 Nikola Buri\' c$^1$,  \thanks{e-mail: buric@phy.bg.ac.yu},Dragana Rankovi\' c$^2$,\\  Kristina Todorovi\' c$^2$
 and  Neboj\v sa Vasovi\' c$^3$\\$1$ Institute of Physics, University of Beograd\\
 PO Box 68, 11080 Beograd-Zemun, Serbia
\\$2$  Department of Physics and
Mathematics,\\ Faculty of Pharmacy, University of Belgrade,\\
Vojvode Stepe 450, Belgrade, Serbia.
\\ $3$ Department of Applied Mathematics, \\
 Faculty of Mining and Geology, University of Belgrade, \\ P.O.Box 162,
 Belgrade, Serbia. }

\maketitle

\begin{abstract}

Mean field approximation of a large collection of FitzHugh-Nagumo
excitable neurons with noise and all-to-all coupling with explicit
time-delays, modelled by $N\gg 1$ stochastic delay-differential
equations is derived. The resulting approximation contains only
two deterministic delay-differential equations but provides
excellent predictions concerning the stability and bifurcations of
the averaged global variables of the exact large system.

\end{abstract}


\newpage

\section{ Introduction}

Small parts of brain cortex may contain thousands of
morphologically and functionally similar interconnected neurons.
Realistic models of  an individual neuron, like Hodgkin-Huxley,
FitzHugh-Nagumo (FN) or Hindmarsh-Rose  to mention only a few
popular examples \cite{Isz}, are given by few-dimensional
nonlinear differential equations. Transport of information between
neurons can be phenomenologically described by time-delayed
inter-neuronal interaction (please see \cite{us1} and the
references therein). It is also well known that neurons {\it in
vivo} function under influences of many sources of
noise\cite{noise-gen}. Considering all mentioned factors it is
clear that a basic, relatively detailed mathematical model of a
small part of realistic cortex should involve an extremely large
system of nonlinear stochastic delay-differential equations
(SDDE). Analyzes of such complex models is impossible without more
or less severe approximations, which should be adopted to
different purposes. It is our goal to study some aspects of an
approximation by only two deterministic delay-differential
equations (DDDE) of
 an example of a complex neuronal system described by
many-component SDDE. We shall see that, although the approximate
model is very simple, the predicted critical parameter values for
the bifurcations and stability of the stationary states are in
excellent quantitative
  agrement with those of the exact complex model within a
  relevant domain of parameters.

Neuronal dynamics with all three factors (large number of units,
delayed interaction and noisy environment ) included has  been
studied much less than the influence of each of the factors
separately \cite{Haken}. Important influence of noise alone on a
single, small number or large clusters of neurons has been studied
a lot in recent years \cite{review}. It is also well known that
time-delay can have important qualitative effects on the stability
of stationary states (please see for example
\cite{us1},\cite{us2}) and synchronization of neuronal dynamics
\cite{dzamala}. Studies of combined effects of noise and
time-delay have mostly, but not entirely (\cite{Hasagawa3}) been
restricted to artificial networks \cite{Blythe} \cite{SunWang} or
small number of neurons (usually two) \cite{us3},\cite{nemci}. An
example of a study of a large collection of noisy realistic
neurons with delayed coupling can be found in \cite{Hasagawa3}
(see also the references therein).

The mean field approach (MFA) is based on a set of approximations
that replace many component system by a simpler system  described
by a small number of (averaged) collective or macroscopic
properties. The mean field approximation has been applied on
systems of excitable neurons with noise but with no time-delay for
example in \cite{Takvel},\cite{Tanabe},\cite{Chaos},\cite{review}.
On the other hand a type of MFA was devised in \cite{Hasagawa1}
and \cite{Hasagawa2} and applied on large clusters of noisy
neurons with time-delayed interaction in \cite{Hasagawa3}.
However, the approximations made in these papers resulted in a
system of equations that is still to large to be analyzed
analytically, so that the approximate system must be studied
numerically. We shall derive an approximate system of only two
DDDE for the dynamics of the mean fields.  Such a simple system
allows analytical treatment of bifurcations and the parameter
domains of stability of the stationary states which turn out to be
in a quite good agrement with the exact complex system.

\section{ The model and its mean field approximation}

We shall study a system  of excitable neurons modelled by the
following set of SDDE:
\begin{eqnarray}
\epsilon dx_i&=&f(x_i,y_i)dt+{c\over N}\sum_{j=1}^N (x_j(t-\tau)-x_i)dt\nonumber\\
dy_i&=&g(x_i,y_i)+\sqrt{2D} dW_i
\end{eqnarray}
with
\begin{eqnarray}
f(x,y)&=&x-x^3/3-y+I\nonumber\\
g(x,y)&=&x+b,
\end{eqnarray}
where $b, I,c,D$ and $\epsilon\ll 1$ are parameters. The formulas
(2) represent one of the common ways of writing the famous
FitzHugh-Nagumo  model \cite{Isz} of the excitable behavior. For
certain parameter values, like $b=1.05,I=0$ to be used throughout
this paper, the ODE given by (2) have stable stationary solution
$(x_0,y_0)$ such that small departures from $(x_0,y_0)$ might lead
to large and long lasting excursions away from $(x_0,y_0)$ which
nevertheless end up on the stable state $(x_0,y_0)$. The type of
excitable behavior epitomized by the FN
 model is called type II \cite{Isz} and is characterized by destabilization
 of the stationary state via the Hopf bifurcation. The variable
 $x$ is called the fast variable (due to $\epsilon \ll 1$) and
 corresponds to the membrane electrical potential. The variable $y$
 is the slow recovery variable and has no direct interpretation.

 Each of  $i=1,2\dots N$ units in (1) is coupled with each other
 unit and with itself. There are two major types of inter-neuronal
 couplings: the chemical and the electrical  synapses.
 Time-delay $\tau$ is important especially in the first type of synapses
 but plays also an important role in the electrical junctions and
 in the transmission of an impulse through the dendrite. In (1) we
 use the electrical coupling with the time-lag and the strength that
 is equal for all pairs of neurons.

 The terms $\sqrt{ 2D} dW_i$ represent  stochastic increments
 of independent Wiener processes, i.e. $dW_i$ satisfy
 \begin{equation}
 E(dW_i)=0,\quad E(dW_idW_j)=\delta_{i,j}dt,
 \end{equation}
 where $E()$ denotes the expectation over many realizations of the
 stochastic process.

{\it Mean field approximation}

 In order to derive the approximate dynamical equations for the  mean
 fields
 \begin{equation}
 X(t)={1\over N}\sum_i^N x_i(t)\equiv <x_i(t)>,\quad Y(t)={1\over N}\sum_i^N
 y_i(t)\equiv <y_i(t)>
 \end{equation}
  of the system (1)
 to be used in this paper we shall first suppose that:
  a) The dynamics is such that the distributions of $x_i$ and $y_i$ are Gaussian
  and b) for large $N$ the average over $N$ of local random variables is given by the expectation with respect to the corresponding distribution,
   i.e. for example ${1\over N}\sum_i^N x_i\approx E(x_i)$, where $E(x_i)$ is the expectation with respect to the
   distribution of $x_i(t)$.  In the limit $N\rightarrow \infty$ the last assumption is expected to become an equality,
   implied by the strong low of large numbers \cite{Arnold}. In the mean field approach it is commonly assumed
    that b) is approximately true even for finite but large
    $N$ despite the nonzero interaction between the local random variables.
    The first assumption should be expected to be true when the noise intensity is small,
    i.e. $D\ll 1$ (see for example \cite{Tanabe},\cite{Chaos}). With these assumptions the system (1) of 2N SDDE can be
reduced to five DDDE for the macroscopic variables $X(t),Y(t)$ and
the second order cumulants. Further assumption concerning the time
scales of first and second order cumulants enables us to derive
the final approximate system of only two DDDE.

Mean field assumption guaranties that global averages, like
$(1/N)\sum_i^N x_i$ of local quantities are equal to the
expectations with respect to distribution of the corresponding
variable $E(x_i)$. Besides the mean values $X(t),Y(t)$ we
introduce deviations from the expectations:
$n_{x_i}(t)=X(t)-x_i(t)$, $n_{y_i}(t)=y_i(t)-Y(t)$. Because of the
assumed Gauss distribution of each variable the first and the
second order cumulants of these deviations are equal to the first
and second order moments ( i.e. to the first and second order
centered moments of the variables $x_i$, etc\dots). Furthermore,
due to the same Gaussian assumption higher order cumulants are
equal to zero, and this enables us to terminate the cumulant
expansion of the dynamical equations. Details of the derivation
are given in the appendix.  The result is a system of five
 deterministic delay-differential equations for the
global variables and global centered moments:
\begin{equation}
s_x=<n^2_{x_i}(t)>,s_y=<n^2_{y_i}(t)>,u=<n_xn_y>.
\end{equation}
 The equations are
\begin{eqnarray}
\epsilon
{dX(t)\over dt}&=&X(t)-X(t)^3/3-s_x(t)X(t)-Y(t)+c(X(t-\tau)-X(t)),\nonumber\\
{dY(t)\over dt}&=&X(t)+b,\nonumber\\
 {\epsilon\over2}
{ds_x(t)\over dt}&=&s_x(t)(1-X(t)^2-s_x(t)-c)-u(t)\nonumber\\
 {1\over2} {ds_y(t)\over dt}&=&u(t)+D,\nonumber\\
 {du(t)\over dt}&=&{u(t)\over
\epsilon}(1-X(t)^2-s_x(t)-c)-{1\over\epsilon}s_y(t)+s_x(t).
\end{eqnarray}

The analogous set of ordinary differential equations  was used to
study the mean field approximation of the stochastic system of N
neurons without delay in \cite{review}. The equations (6) are
delay-differential equations because the original system of
stochastic eq. (1) contains time-delay.

In order to further simplify the approximate system we shall
suppose that relaxation time-scale of the second order moments is
much faster then those of the first order moments. Thus we can
replace in eq. (6) the stationary values of $s_x,s_y$ and $u$
obtained by setting the right hand sides of the last three
equations in (6) equal to zero. As the results we obtain the
following two DDDE:
\begin{eqnarray}
\epsilon {dX(t)\over dt}&=&X(t)-X(t)^3/3-{X(t)\over 2}\left[1-c-X(t)^2+((c-1+X(t)^2)^2+4D)^{1/2}\right]\nonumber\\
&-&Y(t)+c(X(t-\tau)-X(t)),\nonumber\\
{dY(t)\over dt}&=&X(t)+b.\nonumber\\
\end{eqnarray}

\section{ Stability and bifurcations of the stationary state}

Stationary states, their stability and local bifurcations of the
approximate system of DDDE (7) are determined by the standard
procedure. It is remarkable that such a crude approximation
provides relevant information about the exact system.

There is only one stationary state of (7) given by:
\begin{equation}
X(t)\equiv X_0=-b,\quad Y(t)\equiv Y_0={-b\over
2}\left[1+b^2/3+c-(4D+(c+b^2-1)^2)^{1/2}\right].
\end{equation}
Local stability of (8) is determined from the roots of the
characteristic equation. Due to the time-delay the system (7) has
an infinite-dimensional state space, and the characteristic
equation is transcendental with an infinite number of roots. The
characteristic equation is
\begin{eqnarray}
\lambda^2&-&{1\over
2\epsilon}\left[1-c+b^2-(m^2+4D)^{1/2}-{2b^2m\over
(m^2+4D)^{1/2}}\right]\lambda\nonumber\\
&+&{1\over \epsilon}-{c\over \epsilon}\lambda \exp
(-\lambda\tau)=0.
\end{eqnarray}
where $m=c-1+b^2$.

 Bifurcations of the stationary state occur for those values
of the parameters such that any of the infinite number of roots of
(9) has the real part equal to zero \cite{Lunel}. This is possible
only if $\lambda=i\omega$, where $\omega$ can be taken to be
positive. Substitution of $\lambda=i\omega$ in (9) gives
\begin{equation}
\sqrt{2}\omega_{\pm}=\left[ (-k^2+c^2/\epsilon^2+2/\epsilon
\pm(k^2-c^2/\epsilon^2-2/\epsilon)^2-4/\epsilon^2)^{1/2}\right]^{1/2},
\end{equation}
where
$$
k={1\over 2\epsilon}\left[1-c+b^2-(m^2+4D)^{1/2}-{2b^2m\over
(m^2+4D)^{1/2}}\right].
$$
The critical values of the time-lag $\tau$ are related to the
other parameters $c,D$ an $b$ by
\begin{equation}
\tau^j_{c,\pm}=\left[\cos^{-1}(-k\epsilon/c)+2j\pi)\right ]
/\omega_{\pm},\>j=0,1,2\dots
\end{equation}
if
\begin{equation}
{-\omega_{\pm}^2+1/\epsilon\over c\omega_{\pm}/\epsilon}\ge 0
\end{equation}
If (12) is not satisfied then
\begin{equation}
\tau^j_{c,\pm}=\left[-\cos^{-1}(-k\epsilon/c)+(2j+2)\pi\right]/\omega_{\pm},\>j=0,1,2\dots
\end{equation}
\begin{figure}
\includegraphics[height=0.3\textheight,width=0.8\textwidth]{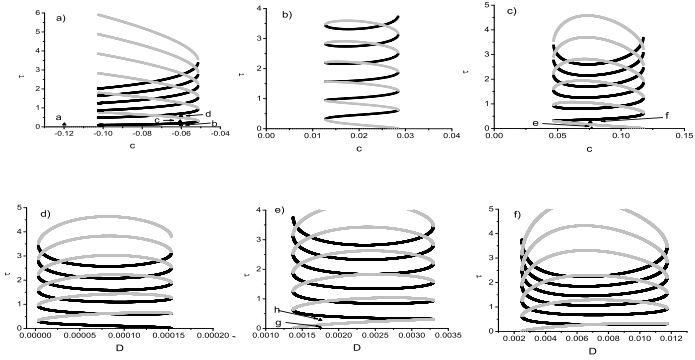}
\caption{Bifurcation curves $(\tau^j_{c,\pm},c)$ (fig. a,b,c) for
fixed $D=0\> (a),$ $D=0.001 \>$ (b), $D=0.003 \>$ (c) and
$(\tau^j_{c,\pm},D)$ (fig. d,e,f)  for fixed $c=-0.05\>$ (a)
$c=0.05 \>$ (b) $c=0.1 \> $. In all figures $b=1.05$. Gray curves
correspond to $\tau^j_{c,-}$ and black curves to $\tau^j_{c,+},\>
$ for $j=0,1,2,3,4,5$.}
\end{figure}
\begin{figure}
\includegraphics[height=0.3\textheight,width=0.8\textwidth]{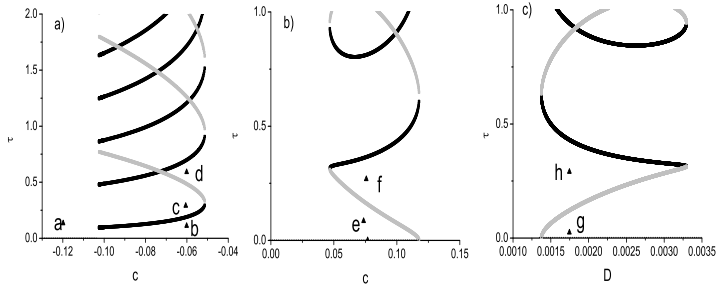}
\caption{Enlarged parts of bifurcation diagrams presented in fig.
1a,c,e with parameter values, indicated by letters: a,b,c,d (fig.
a),e,f (fig. c),g,h (fig. e), that are used for comparison with
the exact system presented in fig. 3. }
\end{figure}

It can be shown by direct substitution that
\begin{equation}
\left({d\Re \lambda\over d\tau}\right)_{\tau=\tau_{c,+}}> 0,\qquad
\left({d\Re \lambda\over d\tau}\right)_{\tau=\tau_{c,+}}< 0,
\end{equation}
so that on the bifurcation curves $\tau^j_{c,+}$ or $\tau^j_{c,+}$
one unstable direction is created or destroyed. Together with the
stability properties for $\tau=0$ the bifurcation curves (11),(13)
and (10) completely solve the problem of stability of the
stationary state. It can be shown by rather lengthy calculations
that the bifurcations for $\tau^j_{c,\pm}$ are the Hopf
supercritical or subcritical bifurcations of the DDDE (7).
\begin{figure}
 \includegraphics[height=0.3\textheight,width=0.9\textwidth]{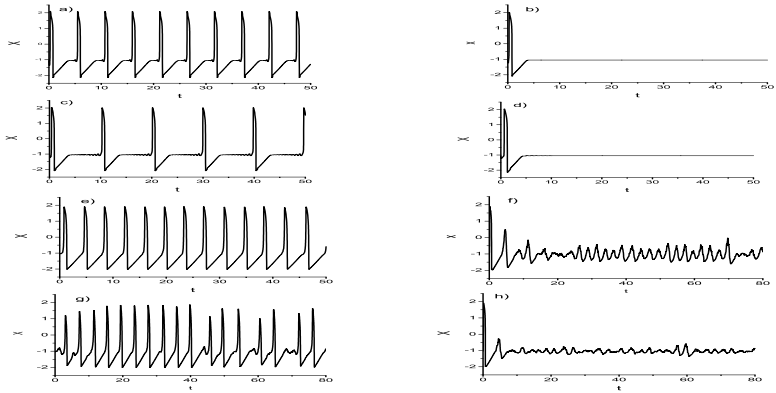}
\caption{Illustrates dynamics of the global variable $X(t)$ for
 the exact system of N=95 units for parameter values corresponding
 to the stable or unstable state of the approximate
 system (7). Parameter values corresponding to a,b,c,d,e,f,g,h, indicated in fig 2, are a) $(c,\tau)=(-0.12,0.14)$;
  b) $(-0.06,0.11)$, c) $(-0.06,0.29)$, d)$(-0.06,0.59)$, e)$(0.07,0.09)$, f)$(0.08,0.27)$, g)$(D,\tau)=(0.002,0.02)$, h)$(0.002,0.29)$. }
\end{figure}

  Bifurcation curves
$\tau^j_{c,\pm}(c)$ for fixed $D,b=1.05$ and $\tau^j_{c,\pm}(D)$
for fixed $c,b=1.05$ are illustrated in figure 1 for different
values of $D$ (fig. 1a,b,c) and $c$ (fig. 1d,e,f). The value
$b=1.05$ renders  the stationary state $X_0,Y_0$ stable and
excitable when $\tau=0$ and $D=0$.

The predictions given by the bifurcation values (11) and (13) of
the system (7) are check versus the numerical solutions of the
exact system (1). To check the approximate predictions of the
bifurcations of stability for the noisy system, i.e. when $D\neq
0$, a proper notion of stochastic bifurcations would be necessary
\cite{Arnold}. Instead we use the sample paths of the SDDE (1)
with large $N$ and for $D\neq 0$ to illustrate that these paths
remain in the vicinity of the stationary solution if the
approximate system's stationary state is stable,  or near a
periodic solution when the state of the approximate system is
unstable.
 Figure 2 presents enlarged parts of bifurcation
diagrams in figure 1, where particular values of the parameters
that correspond either to stable or to unstable stationary state
of (7) are indicated. These parameter values are replaced in the
original system (1) with large $N$ and particular sample paths of
(1) with these parameter values are computed numerically. Time
series of the global variable $X(t)$ along such sample paths are
shown in figure 3, for the system (1) with $N=95$. There is
nothing special with $N=95$ and the same qualitative behavior of
$X(t),Y(t)$ is obtained for any moderately large $N$. It is clear
that when the bifurcation diagrams of the approximate system (7)
predict that the stationary solution is stable ( like in the
cases: b,d,f,h) the sample paths of the exact system display small
stochastic fluctuations around the stationary state. On the other
hand, when the stationary state of (7) is unstable, as shown in
the bifurcation diagrams, the sample paths of the exact system
displays coherent oscillations with large amplitude, indicating
that the exact system has stochastically stable periodic solution.
 The quantitative agrement between the domains of stability in
  the parameters $(D,c,\tau)$ space of the approximate system (7) and the exact system
  (1) is indeed quite
remarkable. It should be expected that such an agrement should be
observed for small values of $D$ since this is one of the
conditions which guaranties that the local random variables have
Gaussian distribution which is one of the asumptions in the
derivation of the mean field equations. It is interesting to
observe the domains of the time-lag where the bifurcation diagrams
in fig. 1 and fig. 2 predict that the non-zero time-lag induces
stabilization of the stationary state. This secession of
oscillations due to the specific non-zero interval of the time-lag
values is correctly predicted for the global variables of the
exact system.

 It should
be stressed that the agrement in the predictions of the
approximate system and the large exact system goes only as far as
the parameter domains of stability are considered. It should be
expected that the predictions of the parameters stability domains
based on (7) should well approximate the parameters stability
domains of the exact system for small values of $D$ since this is
one of the basic assumption in the derivation of the mean field
equations. Also, interaction strength $c$ should be relatively
small in order for the mean field assumption to be valid for
moderately large (but finite) $N$. However, this domain of small
values of $c$ includes interesting bifurcations predicted by (7)
and occurring in (1).
 On the other hand, large values
of $\tau$ induce unstable stationary state and stable oscillatory
behavior in both systems (7) and (1) so formally there is no
restriction on the time-lag $\tau$. We should make clear that it
should not be expected that the values of $X$ and $Y$ for the
deterministic approximate system (7) should reproduce stochastic
orbits $X(t),Y(t)$ for the large exact system or their ensemble
averages. The correspondence between the orbits of the two systems
for the same values of the parameters in different domains is only
qualitative in the sense that they share the same types of
attractors.

\section{ Summary}

We have studied validity of the mean field approximation for the
treatment of stability and bifurcations of the stationary state of
a large collection of FitzHugh-Nagumo excitable neurons with noise
and all-to-all coupling with delays, modelled by $N\gg 1$
stochastic delay-differential equations. Standard assumptions of
the mean field approach are used to derive the system of only two
deterministic delay-differential equations. The stability and
bifurcations of the stationary state of the approximate system can
be studied analytically. The bifurcation curves of the approximate
system give relevant information about the global variables of the
exact large system. For zero and sufficiently
 small noise there is remarkable quantitative agreement of the parameters bifurcation values. On the
 other hand, it should not be expected that the approximation
 gives applicable results when the noise is to large, primarily because
 the assumption about the Gaussian distribution of values of the
 dynamical variables is not valid for large noise.

Using the approximate system it is predicted, and confirmed by
direct numerical simulations on the large exact system, that the
time-lag in a non-zero interval can stabilize the global variables
onto the stationary values even when for zero time-lag the global
variables perform large oscillations.   This is reminiscent of the
phenomenon of the oscillation's death due to the time-delay,
although in this case the relevant dynamics is that of the
averaged global variables and not that of the individual neurons.

We have derived the mean field approximation for the delayed
coupled noisy system using the example of FitzHugh-Nagumo neurons
in the excitable regime. It is expected that the approximations
are equally valid for noisy delayed coupled type I excitable
systems like the Terman-Wang neurons, or for bursting neurons like
the Hidmarch-Rose model. Also the approximation should be
applicable under the same assumptions  for neurons interacting by
delayed chemical rather then electrical coupling.

 {\bf Acknowledgements}
 This work is
partly supported by the Serbian Ministry of Science contract No.
141003. We should like to acknowledge useful comments of the two referees.

\section{Appendix}

The system of equations (1) and (2) can be written in the form:

\begin{eqnarray}
\epsilon dx_i&=&(x-x^3/3-y+I)dt+ c(<x(t-\tau)>-x_i)dt\nonumber\\
dy_i&=&(x+b)dt+\sqrt{2D} dW_i\nonumber
\end{eqnarray}
where:
$$
<x(t-\tau)>={1\over N}\sum_i(x_i(t-\tau)
$$
The bracket $<x>$ is always used to denote the average over the
$N$ units of the local variable $x_i$, which is, by the mean field
assumption, for large $N$ approximately equal to the average over
the assumed Gauss distribution of the corresponding local variable
$x_i$.

Next we introduce deviations from the mean field:
$$
n_{x_i}(t)=<x(t)>-x_i(t),\quad n_{y_i}(t)=<y(t)>-y_i(t).
$$
Deviations will always appear averaged over $N$ i.e. in the form
of $<n_{x}>$ and $<n_{y}>$ so that the index $i$ is in fact
redundant. Correlations between centered moments are defined as
$$
s_x(t)=<n_x^2>,\quad s_y(t)=<n_y^2>,\quad u(t)=<n_xn_y>
$$

Our goal is to derive the equations governing the evolution of the
averages: $X=<x>,Y=<y>,s_x,s_y,u$. Due to the mean field
assumption, these averages can be computed as averages over the
stochastic distributions of the local quantities, which are by
assumption Gaussian. The equation for the derivatives of
$X=<x>,Y=<y>,s_x,s_y,u$ will contain averages of monomials in
local variables of various orders. In order to handel these we
shall need to use the formulas for the cumulant expansions up the
fourth order the local quantities. The general formulas for the
cumulant expansion can be found for example in \cite{Gardiner}.
Due to the assumed Gaussian distribution the third and the fourth
order cumulants (and all of the higher order) are equal to zero,
which will be used to express averages of monomial in local
variables that appear in the evolution equations.

Using the cumulant formulas one computes the following expressions
which will be used to obtain the evolution equations:

From the cumulant $<<x^2y>>=0$ follows
$$
<x_i^2y_i>=Ys_x+YX^2+2Xu.
$$
From the cumulant $<<x^3y>>=0$ follows
$$
<x_i^3y_i>=3s_xu+3X^2u+YX^3+3XYs_x.
$$
Similarly one obtains:
\begin{eqnarray}
<x_i^2>&=&s_x+X^2,\nonumber\\
<x_i^3>&=&X^3+3Xs_x,\nonumber\\
<x_i^4>&=&X^4+6X^2s_x+3s_x^2,\nonumber\\
<x_iy_i>&=&U+XY.
\end{eqnarray}
These expressions provide the necessary ingredients to obtain the
equations (6).

Takeing the average of the equations for $\dot x$ and $\dot y$
gives the first two equations of the system (6). Next consider the
equation for $\dot s_x$.
\begin{eqnarray}
\dot s_x&=&2<X(t)\dot X(t)-X(t)\dot x_i(t)-x_i(t)\dot
X(t)+x_i(t)\dot x_i(t)>\nonumber\\
&=& -2{X(t)\over
\epsilon}[X(t)-X(t)^3/3-X(t)s_x(t)-Y(t)+c(X(t-\tau)-X(t)]\nonumber\\&+&
{2\over
\epsilon}<x_i(t)^2-x_i(t)^4/3-x_i(t)y_i(t)+cx_i(t)X(t-\tau)-cx_i(t)^2>\nonumber\\&=&
{2\over \epsilon} [-X^2(t)s_x(t)+s_x(t)-s_x^2(t)-u(t)-cs_x(y)]
\end{eqnarray}
which is the third equation (6). In the last equality we used the
expressions obtained from  the cumulant formulas.

Equation for $\dot s_y$ is obtained as follows:
$$
\dot s_y=d<Y(t)^2-2Y(t)y_i(t)+y_i(t)^2>/dt=-2Y(t)\dot
Y(t)-d<y_i(t)>/dt.
$$
Using the Ito chain rule this becomes:
\begin{eqnarray}
&-&2Y(t)[X(t)+b]+<2y_(t)dy_i(t)/dt+2D>\nonumber\\&=&-2Y(t)X(t)-2Y(t)b+<2y_i(t)x_i(t)+2y_i(t)b+2y_i(t){\sqrt
2D}
dW_i+2D>\nonumber\\&=&-2Y(t)X(t)-2Y(t)b+2u(t)+2X(t)Y(t)+2Y(t)b+2D\nonumber\\&=&2u(t)+2D,
\end{eqnarray}
which is the forth equation (6).

Similar calculations result in the $\dot u$ equation (6).
\begin{eqnarray}
\dot
u(t)&=&d<X(t)Y(t)-X(t)y_i(t)-Y(t)x_i(t)+x_i(t)y_i(t)>/dt\nonumber\\
&=&-X(t)\dot X(t)-Y(t)\dot X(t)+<y_i(t)\dot x_i(t)>+<\dot y_i(t)
x_i(t)>=\dots\nonumber\\
&=&{1\over
\epsilon}u(t)[1-X^2(t)-s_x(t)-c]-{1\over\epsilon}s_y(t)+s_x(t).
\end{eqnarray}

\end{document}